\documentclass[usenatbib]{mn2e}
\usepackage{graphicx,tabulary,amsmath,amssymb,upgreek,wasysym,subfigure,overpic}
\usepackage[T1]{fontenc}

\numberwithin{equation}{section}
\title[Solenoidal and compressive turbulence]{On the effects of solenoidal and compressive turbulence in prestellar cores}
\author[O. Lomax, A. P. Whitworth, D. A. Hubber]{O. Lomax\thanks{E-mail: oliver.lomax@astro.cf.ac.uk}$^1$, A. P. Whitworth$^1$, D. A. Hubber$^{2,3}$\\
$^1$School of Physics and Astronomy, Cardiff University, Cardiff CF24 3AA, UK\\
$^2$University Observatory, Ludwig-Maximilians-University Munich, Scheinerstr.1, D-81679 Munich, Germany\\
$^3$Excellence Cluster Universe, Boltzmannstr. 2, D-85748 Garching, Germany}

\begin{document}
\pagerange{\pageref{firstpage}--\pageref{lastpage}} \pubyear{2014}
\maketitle
\label{firstpage}

\begin{abstract}
We present the results of an ensemble of SPH simulations that follow the evolution of prestellar cores for $0.2\,{\rm Myr}$. All the cores have the same mass, and start with the same radius, density profile, thermal and turbulent energy. Our purpose is to explore the consequences of varying the fraction of turbulent energy, $\delta_\textsc{sol}$, that is solenoidal, as opposed to compressive; specifically we consider $\delta_\textsc{sol}=1,\,2/3,\,1/3,\,1/9\;{\rm and}\;0$. For each value of $\delta_\textsc{sol}$, we follow ten different realisations of the turbulent velocity field, in order also to have a measure of the stochastic variance blurring any systematic trends. With low $\delta_\textsc{sol}(<\!1/3)$ filament fragmentation dominates and delivers relatively high mass stars. Conversely, with high values of $\delta_\textsc{sol}(>\!1/3)$ disc fragmentation dominates and delivers relatively low mass stars. There are no discernible systematic trends in the multiplicity statistics obtained with different $\delta_\textsc{sol}$.
\end{abstract}


\section{Introduction}%

Understanding the origin of the stellar initial mass function (IMF) \citep[e.g.][]{K01,C03,C05} is one of the main unresolved problems in star formation. Because the physics regulating star formation (i.e. self-gravity, hydrodynamics, radiative transfer, magnetism, etc.) is highly non-linear, general cases can only be studied with numerical simulations. An intrinsic feature of the initial conditions for such simulations is the imposed turbulent velocity field, and since this velocity field is the source of the density fluctuations that spawn protostars, it is important to be clear about how it is defined.

Simulations of star formation usually follow one of two approaches. The first approach involves the simulation of isolated {\it prestellar cores}, i.e. the small ($R\sim\!0.1\,\mathrm{pc}$), dense ($\rho\ga 3\times 10^{-20}\,\mathrm{g\,cm^{-3}}$) clumps of gas with subsonic or mildly transsonic internal turbulence, in which individual stars or small sub-clusters form \citep[e.g.][]{B98, B00, HBB01, MH03, GW04, DCB04a, DCB04b, GWW04, GWW06, WBWNG09, FBCK10, WNWB10, GFBK11, WWG12}. This approach has the advantage that individual simulations can be performed at high resolution with modest computational resource. Consequently good statistics can be obtained by performing multiple realisations. However, it is still important to ensure that the initial conditions mimic reality as closely as is possible  \citep[e.g.][]{LWHSW14}.

The second approach involves the simulation of the much larger, less dense and more massive \emph{molecular clouds} with highly supersonic internal turbulence, in which prestellar cores form \citep[e.g.][]{KHM00,B09a,B12,FK12,B14}. This approach has the advantage that the evolution includes both the formation of cores, and interactions between them, but it is not always feasible to perform more than one realisation, and it remains {to be seen if} the initial conditions {on this scale} are critical.


In this paper we consider the mildly transsonic turbulent velocity fields used to initiate simulations of individual prestellar cores. Specifically, we explore the effect of changing the fraction of turbulent energy, $\delta_\textsc{sol}$, that is solenoidal as opposed to being compressive. Techniques for measuring the ratio of solenoidal to total turbulent energy have recently been developed \citep{BF14}, but have not yet been widely applied. The values of $\delta_\textsc{sol}$ invoked in numerical simulations are seldom explicitly justified. The most common choices are $\delta_\textsc{sol}=2/3$ \citep[thermal mixture, e.g.][]{WWG12,LWHSW14} and $\delta_\textsc{sol}=1$ \citep[purely solenoidal, e.g.][]{B09a,B12,B14}.

Previous numerical work on highly supersonic turbulence on molecular cloud scales \citep[e.g.][]{FK12} indicates that compressive turbulence can deliver star formation rates up to ten times higher than solenoidal turbulence. {Numerical simulations of very massive cores with $M=100\,\mathrm{M_\odot}$ and $R=0.1\,\mathrm{pc}$ \citep{GFBK11} also show that compressive turbulence accelerates the onset of star formation relative to solenoidal turbulence. Here we study this issue on the much smaller scale of the prestellar cores typically seen in nearby star forming regions (e.g. Ophiuchus), where the turbulence is a lot less vigorous.
}


In \S \ref{initial_conditions} we detail the initial conditions used for the simulations. In \S \ref{numerical_method} we describe the numerical method and the constitutive physics. In \S \ref{results} we present the results, and in \S \ref{discussion} we summarise our conclusions.

\section{Initial Conditions}\label{initial_conditions}%

All the simulations presented here start with a spherical core having total mass $M=3\,\mathrm{M_\odot}$, radius $R=3000\,\mathrm{au}$ and non-thermal velocity dispersion $\sigma_\textsc{nt}=0.44\,\mathrm{km\,s^{-1}}$.These values are similar to those of the SM1 core in the Oph-A clump within the L1688 (Ophiuchus) cloud\footnotemark. The initial background temperature is set to $10\,\mathrm{K}$. 

\footnotetext{Using $1.3\,\mathrm{mm}$ dust continuum observations, \citet{MAN98} estimate that SM1 has a mass of $3.2\,\mathrm{M_\odot}$ and an azimuthally averaged full width at half-maximum of $3600\,\mathrm{au}$. \citet{ABMP07} measure the velocity width of the $\mathrm{N_2H^+}$ (1-0) line and estimate that the three-dimensional non-thermal velocity dispersion is $0.45\,\mathrm{km\,s^{-1}}$.}

\subsection{Modified random turbulent velocity field}%

\subsubsection{Standard random turbulent velocity field}\label{SEC:STANDARD}%

Following \citet{LWHSW14}, each core initially has a turbulent velocity field with power spectrum $P_k\propto k^{-4}$ (Burgers turbulence\footnotemark) where $k=8\pi R/\lambda$ is the wavenumber of a velocity mode having wavelength $\lambda$. In three dimensions, each velocity mode is characterised by (i) a wavevector $\boldsymbol{k}=(k_1,k_2,k_3)$; (ii) an amplitude
\begin{equation}\label{EQN:RANDAMPL}
  \boldsymbol{a}(\boldsymbol{k})=\sqrt{P(k)}
    \begin{pmatrix}
      \mathcal{G}_1 \\
      \mathcal{G}_2 \\
      \mathcal{G}_3 
    \end{pmatrix}\,,
\end{equation}
where the $\mathcal{G}_N$ are random variates from a Gaussian distribution (mean $\mu=0$, standard deviation $\sigma=1$); and  (iii) a phase
\begin{equation}\label{EQN:RANDPHASE}
  \boldsymbol{\varphi}(\boldsymbol{k})=2\,\uppi
    \begin{pmatrix}
      \mathcal{U}_1 \\
      \mathcal{U}_2 \\
      \mathcal{U}_3
    \end{pmatrix}\,,
\end{equation}
where the $\mathcal{U}_N$ are random variates from a uniform distribution on the interval [0,1]. Non-zero amplitudes are given to wavevectors with integer components satisfying
\begin{equation}
  \begin{split}
    0 \leq k_3 < k_\textsc{max}\,, & \\
    -k_\textsc{max} \leq k_2 < k_\textsc{max} & \text{\quad if }k_3>0\,,\\
    0 \leq k_2 < k_\textsc{max} & \text{\quad if }k_3=0\,,\\
    -k_\textsc{max} \leq k_1 < k_\textsc{max} & \text{\quad if }k_3>0\text{ or }k_2>0\,,\\
    0 \leq k_1 < k_\textsc{max} & \text{\quad if }k_3=0\text{ and }k_2=0\,.
  \end{split}
\end{equation}
These wavevectors cover all frequencies up to the Nyquist frequency of a grid with $(2\,k_\textsc{max})^3$ uniformly spaced elements.

\footnotetext{{Strictly speaking, a power law exponent of -4 (Burgers turbulence) is only appropriate for highly supersonic turbulence \citep[see][]{F13}. At sonic and transonic speeds, the exponent is likely to be between -4 and -11/3 (Kolmogorov turbulence). However, with both exponents, the turbulent energy is strongly concentrated at the longest wavelengths, and so the precise choice of exponent is not critical.}}

\subsubsection{Modifications to the longest wavelength modes}\label{SEC:MODIFICATION}%

The longest-wavelength velocity modes, those corresponding to the scale of the core, are then modified so that they generate radial excursions (either contraction or expansion) relative to the centre of the core, and rotation about the centre of the core. This is achieved by revising the amplitudes of the modes $\boldsymbol{k_1}=(1,0,0)$, $\boldsymbol{k_2}=(0,1,0)$ and $\boldsymbol{k_3}=(0,0,1)$ to
\begin{equation}
  \boldsymbol{a}(\boldsymbol{k_1})=
  \begin{pmatrix}
    \mathcal{G}_1 \\
    \mathcal{G}_6 \\
    -\mathcal{G}_5
  \end{pmatrix},\,
  \boldsymbol{a}(\boldsymbol{k_2})=
  \begin{pmatrix}
    -\mathcal{G}_6 \\
    \mathcal{G}_2 \\
    \mathcal{G}_4
  \end{pmatrix},\,
  \boldsymbol{a}(\boldsymbol{k_3})=
  \begin{pmatrix}
    \mathcal{G}_5 \\
    -\mathcal{G}_4 \\
    \mathcal{G}_3
  \end{pmatrix},
\end{equation}
and their phases to
\begin{equation}
  \boldsymbol{\varphi}(\boldsymbol{k_1})=\boldsymbol{\varphi}(\boldsymbol{k_2})=\boldsymbol{\varphi}(\boldsymbol{k_3})=
  \begin{pmatrix}
    \uppi/2 \\
    \uppi/2 \\
    \uppi/2 \\
  \end{pmatrix}\,.
\end{equation}
With this procedure, $\mathcal{G}_1$, $\mathcal{G}_2$ and $\mathcal{G}_3$ determine the amount of global compression (expansion) towards (away from) the centre of the core; and $\mathcal{G}_4$, $\mathcal{G}_5$ and $\mathcal{G}_6$ determine the amount of global rotation about the centre of the core. This adjustment is made because, if a newly-formed core is undergoing global contraction or expansion, these motions are likely to be focussed on the centre of the core. Similarly, if a newly-formed core is undergoing global rotation, these motions are likely to be around the centre of the core. All other velocity modes, up to $k_\textsc{max}=64$ retain their random phases, as generated by Eqns. (\ref{EQN:RANDAMPL}) and (\ref{EQN:RANDPHASE}), and therefore represent internal random turbulence.

\subsubsection{Helmholtz decomposition}\label{SEC:HELMHOLTZ}%

Helmholtz's theorem states that a vector field can be expressed as the sum of a compressive (curl-free) vector field and a solenoidal (divergence-free) vector field. For a velocity mode with wavevector $\boldsymbol{k}$ and amplitude $\boldsymbol{a}(\boldsymbol{k})$, the longitudinal component of the amplitude,
\begin{equation}
  \boldsymbol{a_\textbf{\textsc{l}}}(\boldsymbol{k})=\boldsymbol{k}(\boldsymbol{k}\cdot\boldsymbol{a}(\boldsymbol{k}))\,,
\end{equation}
contributes to the compressive field $\boldsymbol{v_\textbf{\textsc{c}}}(\boldsymbol{x})$, and the transverse component of the amplitude,
\begin{equation}
  \boldsymbol{a_\textbf{\textsc{t}}}(\boldsymbol{k})=\boldsymbol{a}(\boldsymbol{k})-\boldsymbol{a_\textbf{\textsc{l}}}(\boldsymbol{k})\,,
\end{equation}
contributes to the solenoidal field $\boldsymbol{v_\textbf{\textsc{s}}}(\boldsymbol{x})$. These components can be summed as wave amplitudes or in real space,
\begin{equation}
  \begin{split}
    \boldsymbol{a}(\boldsymbol{k})&=\boldsymbol{a_\textbf{\textsc{l}}}(\boldsymbol{k})+\boldsymbol{a_\textbf{\textsc{t}}}(\boldsymbol{k})\,,\\
    \boldsymbol{v}(\boldsymbol{x})&=\boldsymbol{v_\textbf{\textsc{c}}}(\boldsymbol{x})+\boldsymbol{v_\textbf{\textsc{s}}}(\boldsymbol{x})\,,
  \end{split}
\end{equation}
to retrieve the total values. Note that $\boldsymbol{a_\textbf{\textsc{l}}}(\boldsymbol{k})$ is the component of $\boldsymbol{a}(\boldsymbol{k})$ parallel to $\boldsymbol{k}$ and $\boldsymbol{a_\textbf{\textsc{t}}}(\boldsymbol{k})$ is the perpendicular component. In three dimensions -- and assuming that the $\hat{\boldsymbol{a}}(\boldsymbol{k})$ are distributed isotropically -- $\boldsymbol{v_\textbf{\textsc{s}}}(\boldsymbol{x})$ has on average twice the kinetic energy of $\boldsymbol{v_\textbf{\textsc{c}}}(\boldsymbol{x})$. {This is because transverse waves have two degrees of freedom whereas longitudinal waves only have one \citep[see][]{FKS08}. Helmholtz decomposion has also been used in other astrophysical simulations \citep[e.g.][]{SFHK09,FRKS10,GFBK11}.}

For an arbitrary velocity field $\boldsymbol{v}(\boldsymbol{x})$, we can alter the fraction of solenoidal kinetic energy by decomposing and then reconstituting the amplitudes of each velocity mode. Here, we define five sets of modified amplitudes: 
\begin{equation}
  \begin{split}
    \boldsymbol{a_1}(\boldsymbol{k})&=\boldsymbol{a_\textbf{\textsc{t}}}(\boldsymbol{k})\,,\\
    \boldsymbol{a_2}(\boldsymbol{k})&=\boldsymbol{a_\textbf{\textsc{l}}}(\boldsymbol{k})+\boldsymbol{a_\textbf{\textsc{t}}}(\boldsymbol{k})\,,\\
    \boldsymbol{a_3}(\boldsymbol{k})&=2\,\boldsymbol{a_\textbf{\textsc{l}}}(\boldsymbol{k})+\boldsymbol{a_\textbf{\textsc{t}}}(\boldsymbol{k})\,,\\
    \boldsymbol{a_4}(\boldsymbol{k})&=4\,\boldsymbol{a_\textbf{\textsc{l}}}(\boldsymbol{k})+\boldsymbol{a_\textbf{\textsc{t}}}(\boldsymbol{k})\,,\\
    \boldsymbol{a_5}(\boldsymbol{k})&=\boldsymbol{a_\textbf{\textsc{l}}}(\boldsymbol{k})\,.
  \end{split}
\end{equation}
The average fractions of kinetic energy in solenoidal modes for the corresponding velocity fields, $\boldsymbol{v_1}(\boldsymbol{x})$, $\boldsymbol{v_2}(\boldsymbol{x})$, $\boldsymbol{v_3}(\boldsymbol{x})$, $\boldsymbol{v_4}(\boldsymbol{x})$ and $\boldsymbol{v_5}(\boldsymbol{x})$, are respectively $\delta_\textsc{sol}=1,\,2/3,\,1/3,\,1/9\;{\rm and}\;0$.

\subsubsection{Particle velocities}

Having defined all the velocity modes, we use the fast Fourier transform library FFTW \citep{FFTW05} with $k_\textsc{max}=64$ to compute a gridded velocity field $\boldsymbol{v}(\boldsymbol{x})$ on $-2R\leq x_1,x_2,x_3\leq +2R$, and the velocities from the central eighth of the volume are then mapped onto the SPH particles.

\subsection{Density profile}

Many observations \citep[e.g.][]{ALL01,HWL01,KWA05,LMR08,RAP13} suggest that the critical Bonnor-Ebert sphere provides a good fit to the column-density profile of a prestellar core, {\it even if the core is not in hydrostatic equilibrium}. We therefore set the core density profile to $\rho(\xi)=\rho_\textsc{c}\,\mathrm{e}^{-\psi(\xi)}$, where $\rho_\textsc{c}$ is the central density, $\psi(\xi)$ is the Isothermal Function and $\xi$ is the dimensionless radius, i.e. $\xi=6.451 (r/3000\,{\rm au})$.

\subsection{Parameter space}%

Using the procedures described in Sections \ref{SEC:STANDARD} and \ref{SEC:MODIFICATION}, we generate ten different initial velocity fields, by invoking ten different random seeds, 
\begin{equation}
{\cal I}_\textsc{seed}=1,\,2,\,3,\,4,\,5,\,6,\,7,\,8,\,9\;{\rm and}\;10\,.
\end{equation}
Then, using the procedures described in Section \ref{SEC:HELMHOLTZ}, we convert each of these velocity fields into five velocity fields with different fractions of solenoidal kinetic energy
\begin{equation}
\delta_\textsc{sol}=1,\,2/3,\,1/3,\,1/9\;{\rm and}\;0\,.
\end{equation} 
We therefore have a total of 50 initial velocity fields, corresponding to all possible combinations of the ten different random seeds, ${\cal I}_\textsc{seed}$, and the five different fractions, $\delta_\textsc{sol}$.

\section{Numerical Method}\label{numerical_method}%

Core evolution is simulated using the \textsc{seren} $\nabla h$-SPH code \citep{HBMW11}, with $\eta = 1.2$ (so a particle typically has $57$ neighbours). Gravitational forces are computed using a tree, and the \citet{MM97} formulation of time dependent artificial viscosity is invoked. In all simulations, the SPH particles have mass $m_\textsc{sph}=10^{-5}\,\mathrm{M}_{\odot}$, so the opacity limit ($\sim\!\!3\times10^{-3}\,\mathrm{M}_{\odot}$) is resolved with $\sim\!\!300$ particles. Gravitationally bound regions with density higher than $\rho_\textsc{sink}=10^{-9}\,\mathrm{g}\,\mathrm{cm}^{-3}$ are replaced with sink particles \citep{HWW13}. Sink particles have radius $r_\textsc{sink}\simeq0.2\,\mathrm{au}$, corresponding to the smoothing length of an SPH particle with density equal to $\rho_\textsc{sink}$. The equation of state and the energy equation are treated with the algorithm described in \citet{SWBG07}.

Radiative feedback from sinks is also included. Each sink has a variable luminosity which follows the episodic accretion model described in \citet{SWH11} \citep[also used in][]{SWH12,LWHSW14}. In this model, highly luminous, short-lived accretion episodes are separated by $\sim\!10^4\,\mathrm{yrs}$ of low-luminosity quiescent accretion (during which matter collects in the inner accretion disc until it is hot enough to become thermally ionised and couple to the magnetic field; then the Magneto-Rotational Instability delivers efficient outward angular momentum transport and the matter is dumped onto the star).

\section{Results}\label{results}%

\begin{figure}
\centering
\includegraphics[width=\columnwidth]{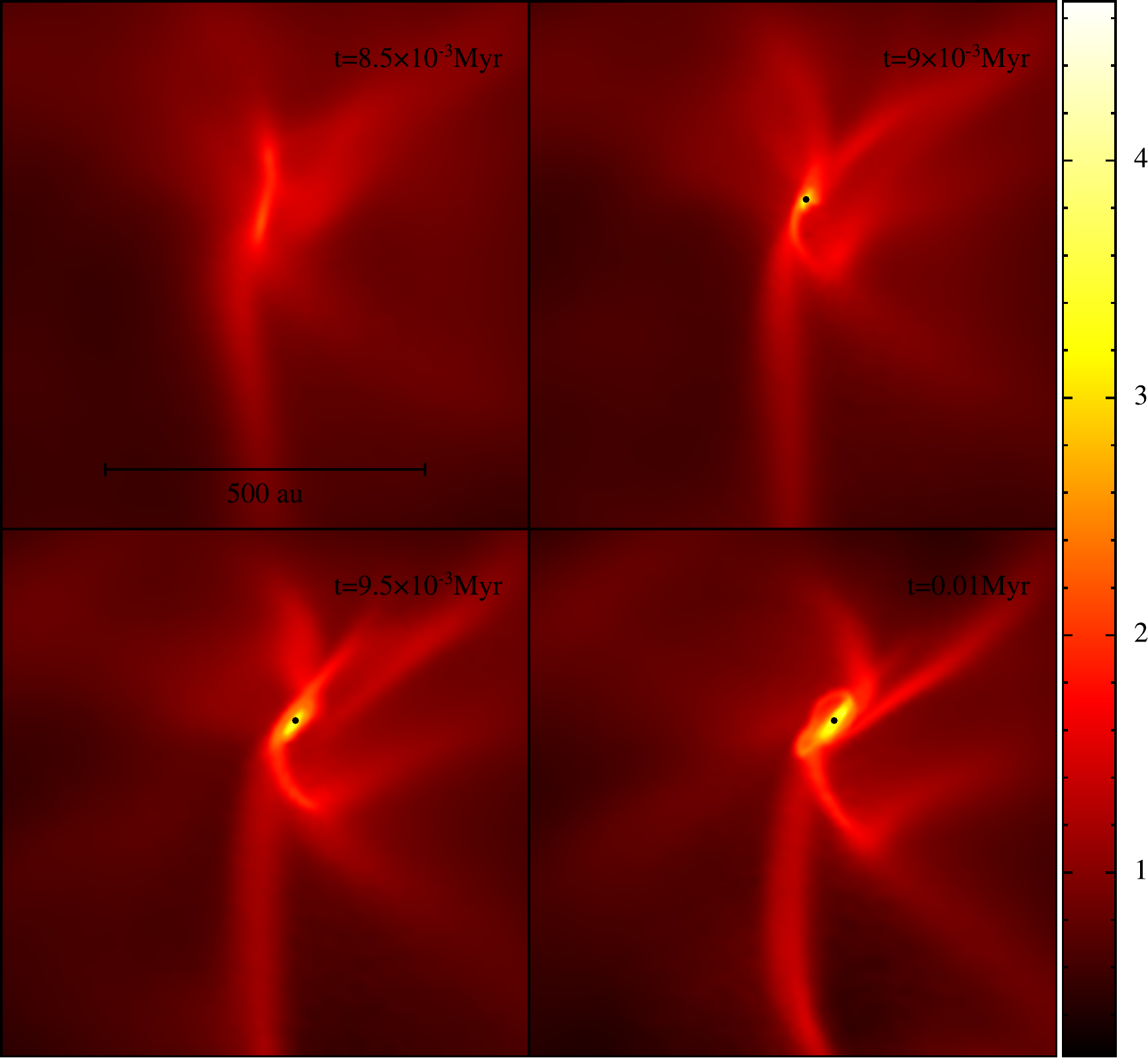}
\caption{False-colour column-density images on the central $820\,{\rm au}$ by $820\,{\rm au}$ of the $(x,y)$-plane, from the simulation with ${\cal I}_\textsc{seed}=3$ and $\delta_\textsc{sol}=2/3$, at times $t=0.85,0.90,0.95\text{ and }1.00\times10^4\,\mathrm{yrs}$. The colour scale gives the logarithmic column density in units of $\mathrm{g\,cm^{-2}}$. Sink particles are represented by black dots. This is an example of filament fragmentation, where the filaments serve to deliver matter from the periphery of the core into the centre. Further evolution of this case is shown in Fig. \ref{disc_1}\,.}
\label{frag_1}
\end{figure}

\begin{figure}
\centering
\includegraphics[width=\columnwidth]{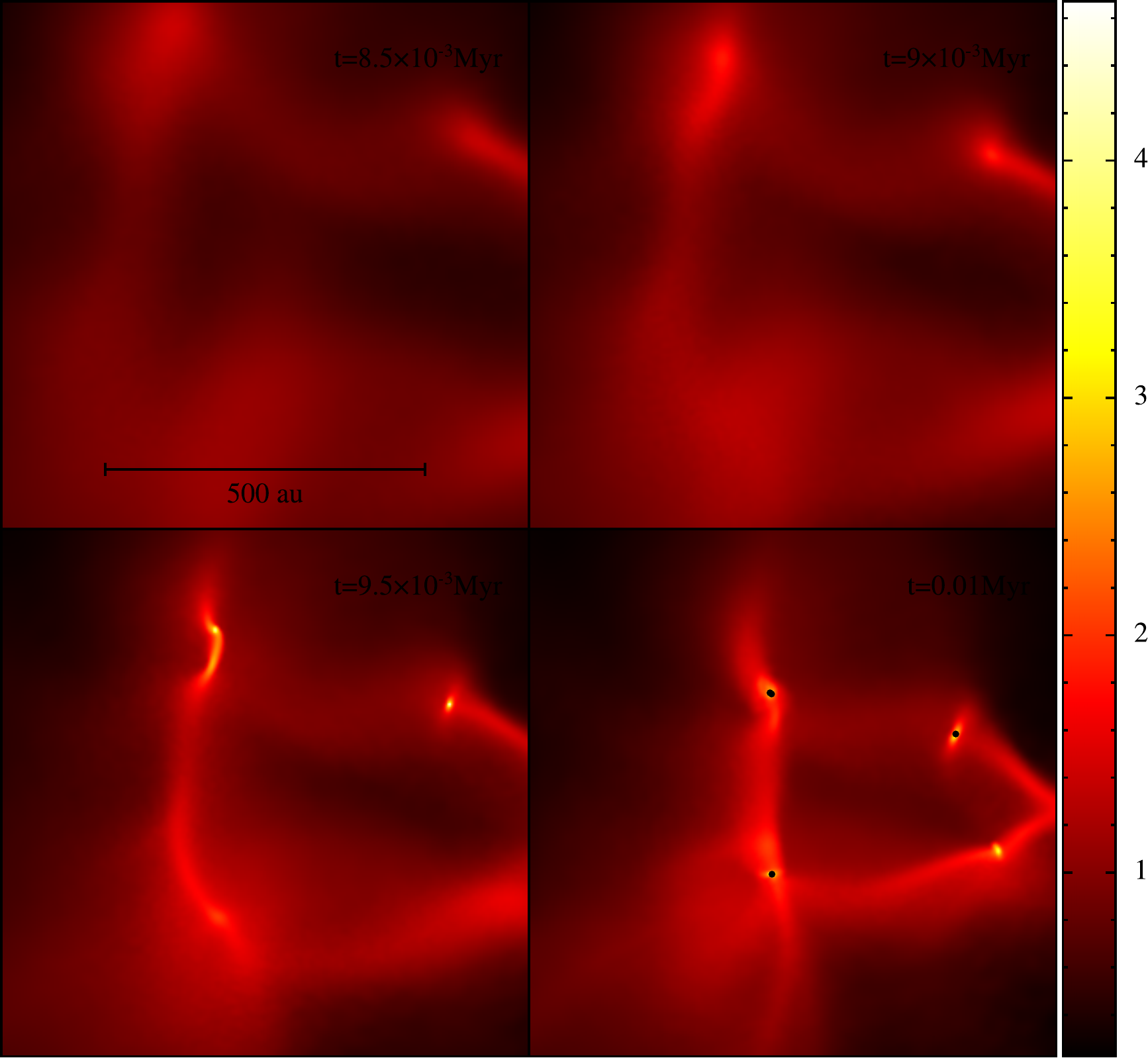}
\caption{False-colour column-density images on the central $820\,{\rm au}$ by $820\,{\rm au}$ of the $(x,y)$-plane, from the simulation with $\mathcal{I}_\textsc{seed}=3$ and $\delta_\textsc{sol}=0$, at times $t=0.85,0.90,0.95\text{ and }1.00\times10^4\,\mathrm{yrs}$. The colour scale gives the logarithmic column density in units of $\mathrm{g\,cm^{-2}}$. Sink particles are represented by black dots. This is an example of filament fragmentation, where the individual filaments fragment independently to produce an ensemble of stars. Further evolution of this case is shown in Fig. \ref{mont_5}.}
\label{frag_2}
\end{figure}

\begin{figure}
\includegraphics[width=\columnwidth]{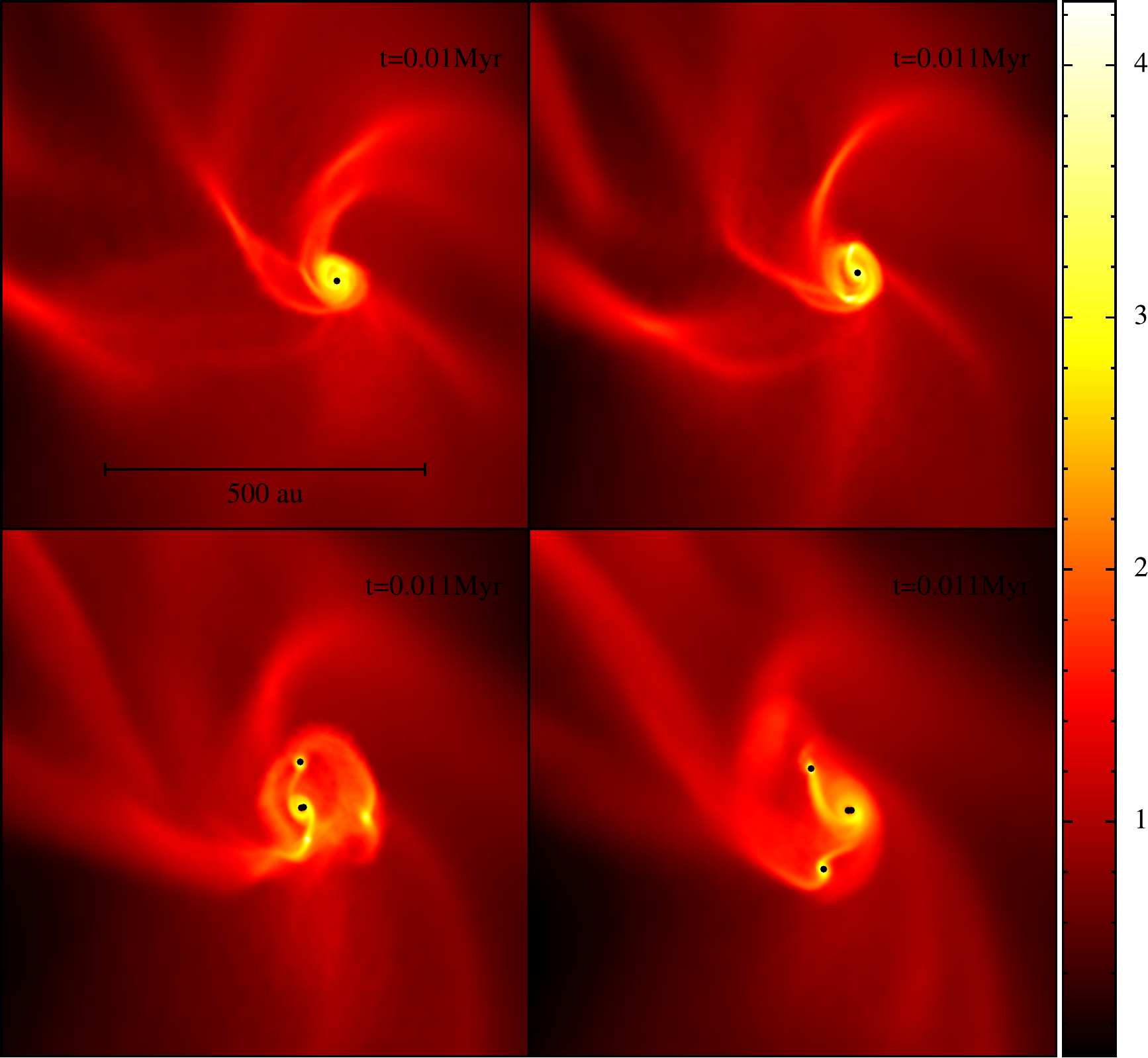}
\caption{False-colour column-density images on the central $820\,{\rm au}$ by $820\,{\rm au}$ of the $(x,z)$-plane, from the simulation with $\mathcal{I}_\textsc{seed}=3$ and $\delta_\textsc{sol}=2/3$, at times $t=1.00,1.05,1.10\text{ and }1.15\times10^4\,\mathrm{yrs}$. The colour scale gives the logarithmic column density in units of $\mathrm{g\,cm^{-2}}$. Sink particles are represented by black dots. This is an example of disc fragmentation. Further evolution of this case is shown in Fig. \ref{mont_2}\,.}
\label{disc_1}
\end{figure}

Each core is evolved for $0.2\,\mathrm{Myr}$. This is roughly the predicted time between core-core collisions in Ophiuchus \citep{ABMP07}. The simulations do not include any mechanical feedback from sinks (e.g. outflows), so the star formation efficiency is high, roughly $0.8<\eta<1.0$. Previous work {\citep[e.g.][]{MM00,FSBK14} demonstrates that outflows and jets can reduce star formation efficiency signficantly.}

\subsection{Modes of fragmentation}%

As a core collapses, turbulence and self-gravity organise the matter into filaments. These filaments usually behave in one of two ways: (i) they feed material from the periphery of the core into its centre, forming a central star, or (ii) they fragment independently to form multiple stars, which then congregate in a small cluster near the centre of the core. An example of case (i) is shown in Fig. \ref{frag_1} and an example of case (ii) is shown in Fig. \ref{frag_2}. In the sequel we refer to this initial mode of star formation as \emph{filament fragmentation}.

Once filament fragmentation has occurred, the remaining matter in the core envelope tries to accrete onto the star or cluster near the centre. If this matter has sufficient angular momentum, it forms circumstellar or circumsystem discs. Discs which are sufficiently massive \citep{T64} and are able to cool sufficiently fast \citep{Gam01} fragment to produce additional stars \citep[e.g.][]{SW08,SW09a,SW09b,SMWA11}. An example of this process is shown in Fig. \ref{disc_1}. In the sequel we refer to this second mode of star formation as \emph{disc fragmentation}.

\subsection{Influence of $\delta_\textsc{sol}$ on the dominant mode of fragmentation}\label{modes_of_frag}%

Table \ref{frag_table} lists the number of stars that form by filament fragmentation, $N_\textsc{ff}$, and the number that form by disc fragmentation, $N_\textsc{df}$ The distinction between the two modes is made by inspecting the simulation frames by eye. This table highlights the need to invoke multiple realisations with different random seeds, since, with a given $\delta_\textsc{sol}$, the results can vary dramatically with $\mathcal{I}_\textsc{seed}$. For example, with $\delta_\textsc{sol}=1/9$, only one star forms when $\mathcal{I}_\textsc{seed}=5$, whereas twelve stars form when $\mathcal{I}_\textsc{seed}=10$. This is because most of the turbulent energy is in large-scale modes which are defined by only a few wavevectors, and so the outcome is very sensitive to the random amplitudes of these modes.

\begin{figure}
  \includegraphics[width=\columnwidth]{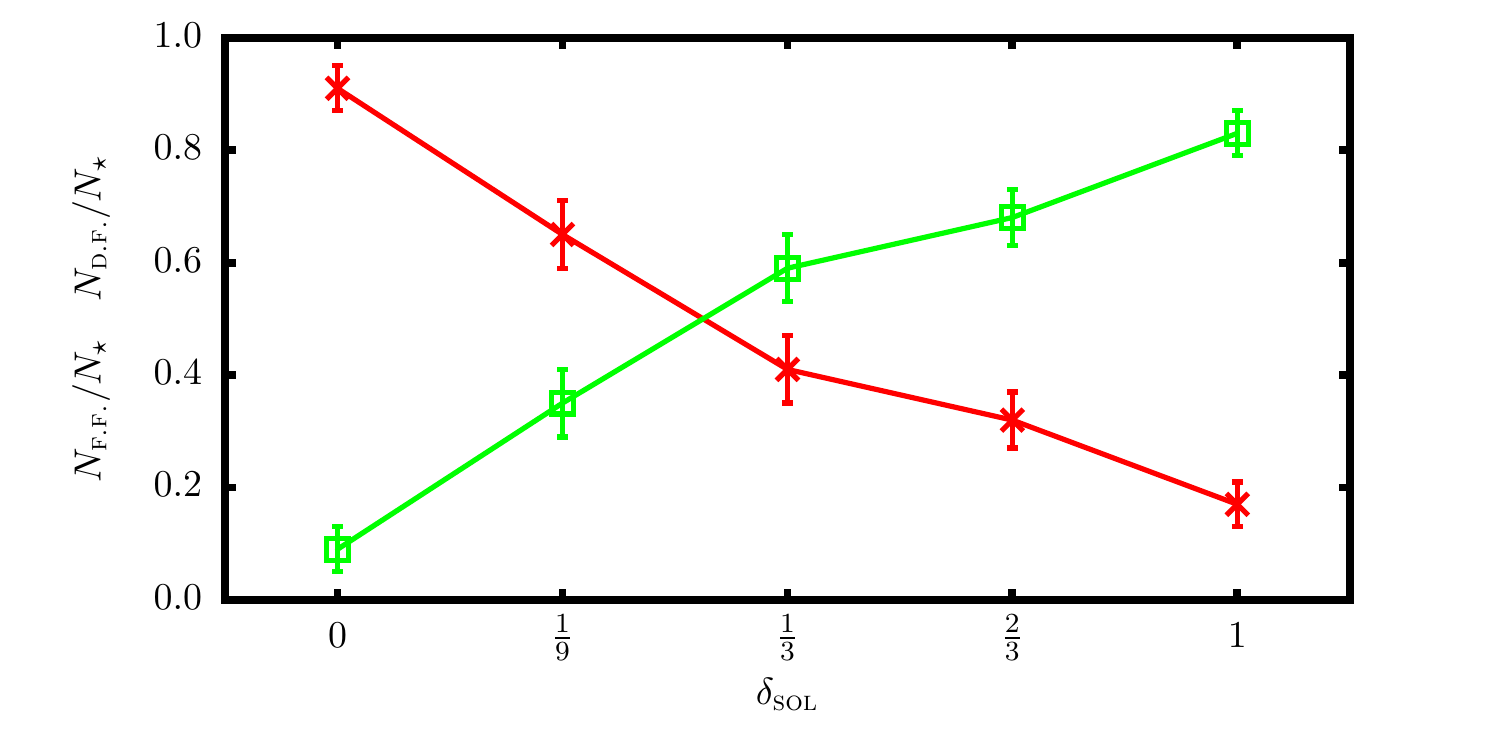}
  \caption{The fraction of stars formed by filament fragmentation (red $\times$s) and disc fragmentation (green $\Box$s) for different values of $\delta_\textsc{sol}$, averaged over all values of $\mathcal{I}_\textsc{seed}$. The error bars give the Poisson counting uncertainties.}
  \label{ff_df}
\end{figure}

\begin{figure*}
  \centering
  \subfigure[$\delta_\textsc{sol}=0$]{\label{mont_5}\includegraphics[width=0.33\textwidth]{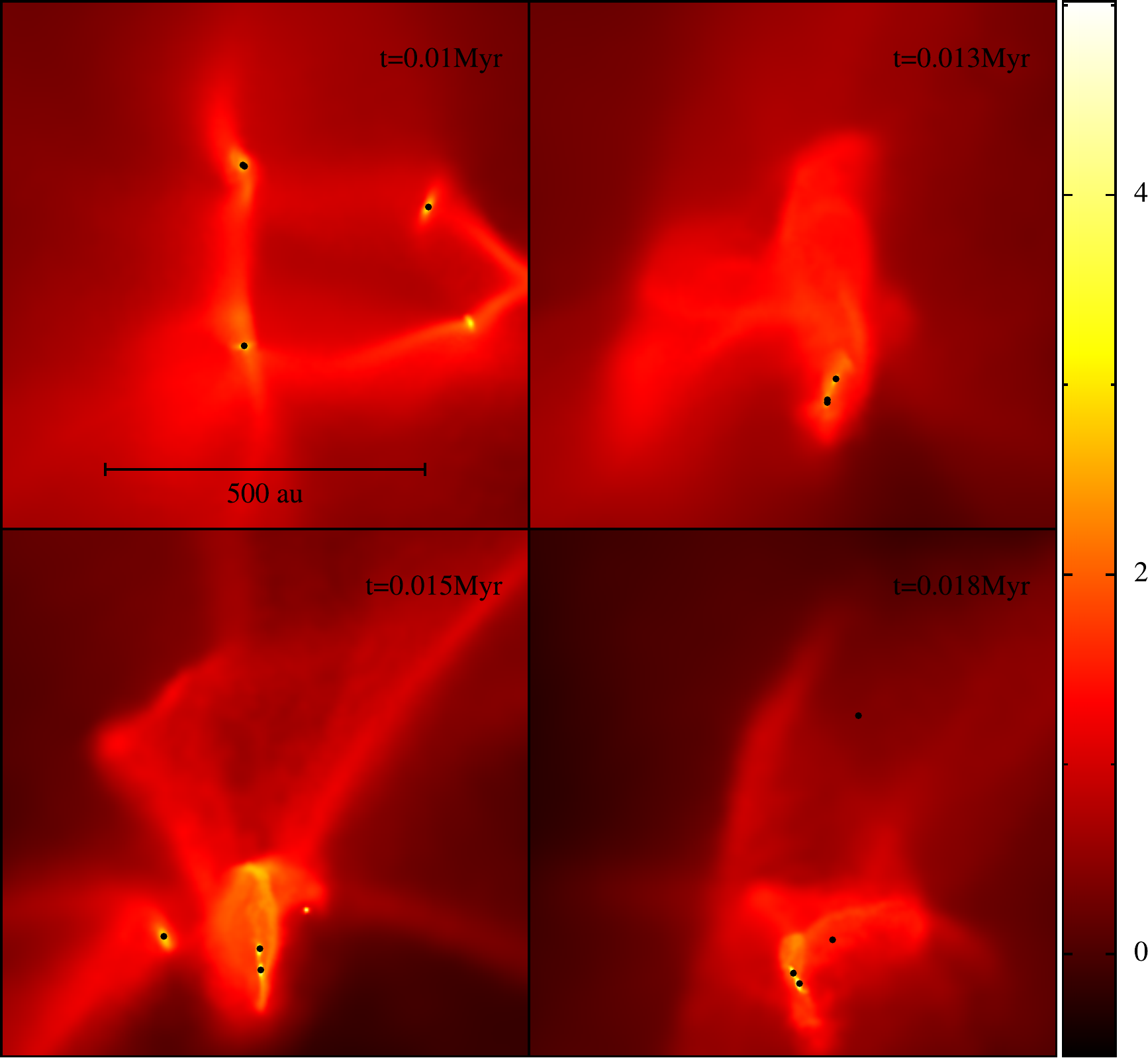}}
  \subfigure[$\delta_\textsc{sol}=1/9$]{\label{mont_4}\includegraphics[width=0.33\textwidth]{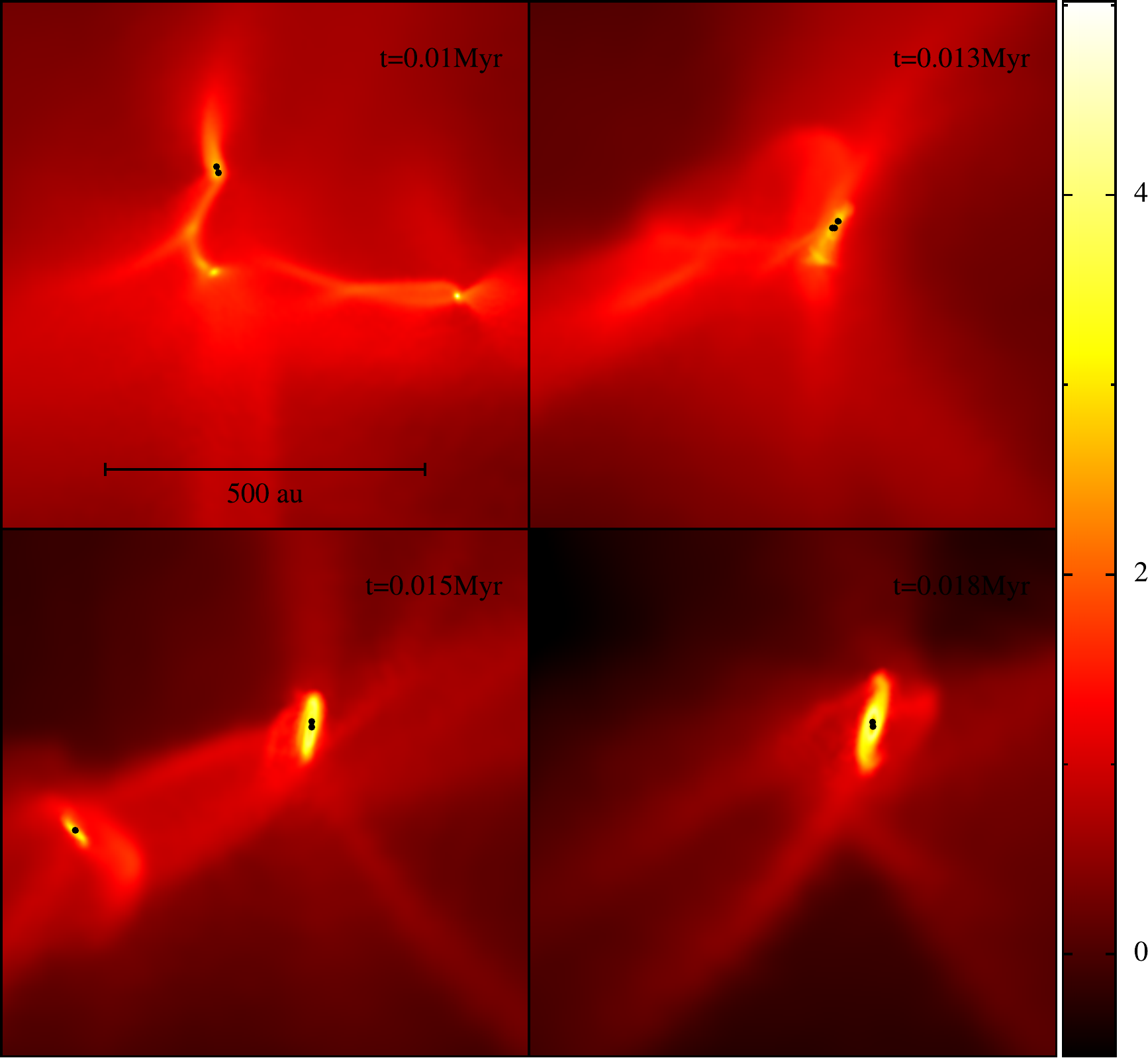}}
  \subfigure[$\delta_\textsc{sol}=1/3$]{\label{mont_3}\includegraphics[width=0.33\textwidth]{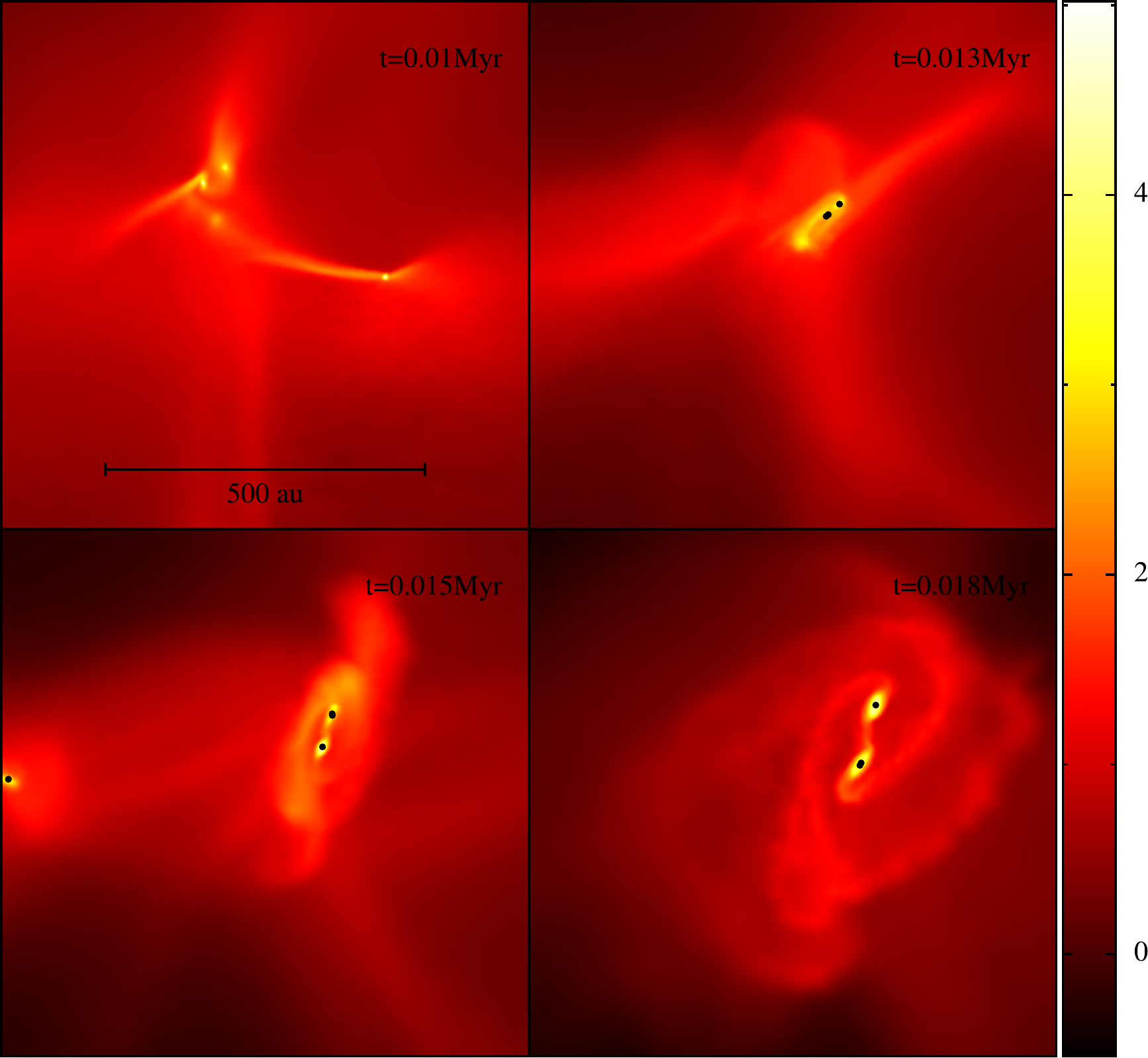}}
  \subfigure[$\delta_\textsc{sol}=2/3$]{\label{mont_2}\includegraphics[width=0.33\textwidth]{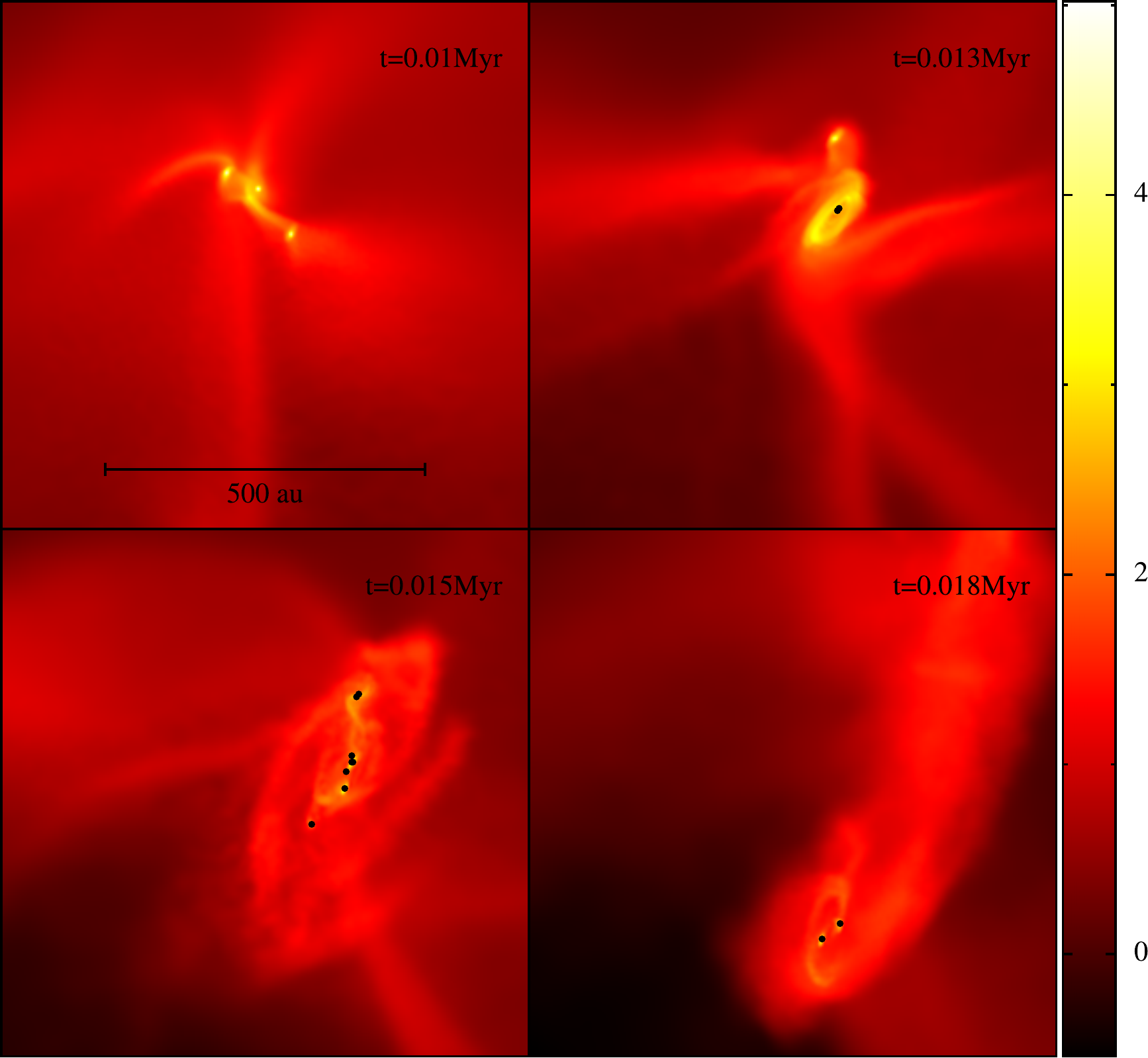}}
  \subfigure[$\delta_\textsc{sol}=1$]{\label{mont_1}\includegraphics[width=0.33\textwidth]{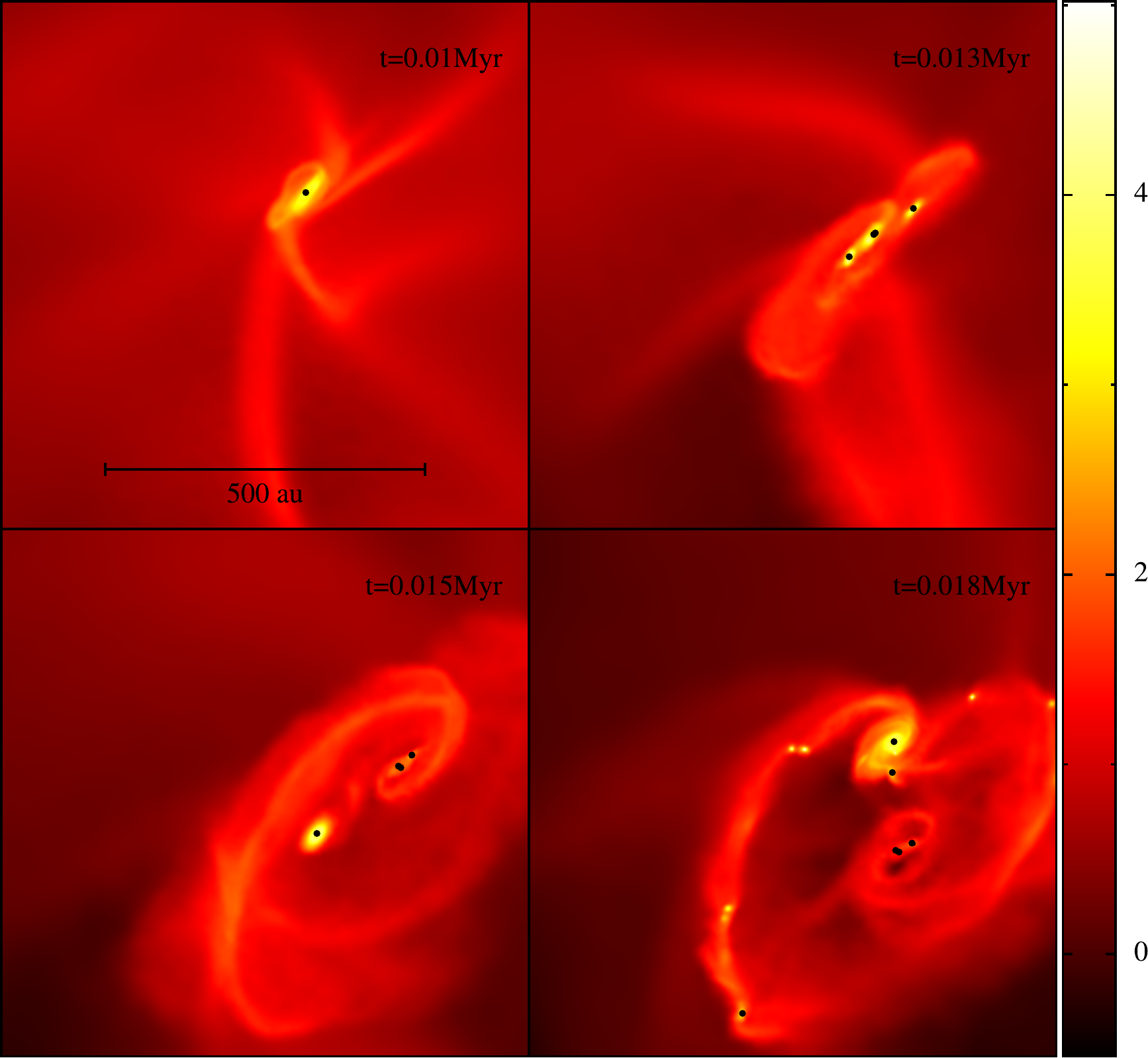}}
  \caption{False-colour column-density images on the central $820\,{\rm au}$ by $820\,{\rm au}$ of the $(x,y)$-plane, from the simulations with $\mathcal{I}_\textsc{seed}=3$ and different values of $\delta_\textsc{sol}$, at times $t=1.00,1.25,1.50\text{ and }1.75\times10^4\,\mathrm{yrs}$. The colour scale gives the logarithmic column density in units of $\mathrm{g\,cm^{-2}}$. Sink particles are represented by black dots.}
  \label{montage}
\end{figure*}

Fig. \ref{montage} shows the early stages of star formation with $\mathcal{I}_\textsc{seed}=3$ and different values of $\delta_\textsc{sol}$.  In Fig. \ref{mont_5}, where $\delta_\textsc{sol}=0$, seven stars form by filament fragmentation, and then a single star by disc fragmentation. In Fig. \ref{mont_1}, where $\delta_\textsc{sol}=1$, a single star forms by filament fragmentation, then thirteen by disc fragmentation. Figs. \ref{mont_4}, \ref{mont_3} and \ref{mont_2} show how the number of stars formed by filament fragmentation tends to decrease, and the number of stars formed by disc fragmentation to increase, with increasing $\delta_\textsc{sol}$. This trend is seen more clearly in Fig. \ref{ff_df} where the results are averaged over all values of $\mathcal{I}_\textsc{seed}$. The average fraction of stars formed from a core by filament fragmentation decreases monotonically from $\sim\!0.9$ when $\delta_\textsc{sol}=0$ to $\sim\!0.2$ when $\delta_\textsc{sol}=1$. Conversely, the average fraction of stars formed from a core by disc fragmentation increases monotonically from $\sim\!0.1$ when $\delta_\textsc{sol}=0$ to $\sim\!0.8$ when $\delta_\textsc{sol}=1$\,. This is because predominantly compressive fields (low $\delta_\textsc{sol}$) are characterised by shocks, and these are conducive to filament formation. Conversely, predominantly solenoidal fields (high $\delta_\textsc{sol}$) are characterised by shearing motions, and these generate the angular momentum required for the formation of discs.

Table \ref{frag_table} also indicates that the total number of stars formed per core increases slightly with increasing $\delta_\textsc{sol}$. When $\delta_\textsc{sol}=0$, a core spawns on average $\sim\!5$ stars; when $\delta_\textsc{sol}=1$, a core spawns on average $\sim\!8$ stars.

{\footnotesize
\begin{table*}
\centering
\resizebox{\textwidth}{!}{
\begin{tabular}{|c|cc|cc|cc|cc|cc|}\hline
&&&&&&&&&&\\
& \multicolumn{2}{c|}{$\delta_\textsc{sol}=0$} & \multicolumn{2}{c|}{$\delta_\textsc{sol}=1/9$} & \multicolumn{2}{c|}{$\delta_\textsc{sol}=1/3$} & \multicolumn{2}{c|}{$\delta_\textsc{sol}=2/3$} & \multicolumn{2}{c|}{$\delta_\textsc{sol}=1$} \\
$\mathcal{I}_\textsc{seed}$ & $N_\textsc{ff}$ & $N_\textsc{df}$ & $N_\textsc{ff}$ & $N_\textsc{df}$ & $N_\textsc{ff}$ & $N_\textsc{df}$ & $N_\textsc{ff}$ & $N_\textsc{df}$ & $N_\textsc{ff}$ & $N_\textsc{df}$ \\
&&&&&&&&&&\\
1 & 6 & 1 & 5 & 1 & 2 & 8 & 3 & 6 & 2 & 6 \\
2 & 4 & 0 & 2 & 4 & 4 & 0 & 4 & 0 & 1 & 12 \\
3 & 7 & 1 & 6 & 0 & 3 & 1 & 4 & 10 & 1 & 13 \\
4 & 5 & 0 & 6 & 1 & 3 & 1 & 2 & 6 & 1 & 3 \\
5 & 5 & 0 & 1 & 0 & 1 & 7 & 3 & 11 & 2 & 6 \\
6 & 5 & 0 & 4 & 1 & 5 & 1 & 2 & 3 & 2 & 1 \\
7 & 6 & 1 & 6 & 1 & 4 & 6 & 4 & 4 & 1 & 9 \\
8 & 3 & 0 & 2 & 2 & 2 & 2 & 2 & 2 & 2 & 4 \\
9 & 4 & 2 & 4 & 4 & 1 & 8 & 2 & 9 & 1 & 8 \\
10 & 4 & 0 & 4 & 8 & 2 & 5 & 2 & 9 & 1 & 5 \\
&&&&&&&&&&\\
Total & $49\pm7$ & $5\pm2$ & $40\pm6$ & $22\pm5$ & $27\pm5$ & $39\pm6$ & $28\pm5$ & $60\pm8$ & $14\pm4$ & $67\pm8$ \\
Fraction & $0.91\pm0.04$ & $0.09\pm0.04$ & $0.65\pm0.06$ & $0.35\pm0.06$ & $0.41\pm0.06$ & $0.59\pm0.06$ & $0.32\pm0.05$ & $0.68\pm0.05$ & $0.17\pm0.04$ & $0.83\pm0.04$ \\ 
&&&&&&&&&&\\\hline
\end{tabular}
}
\caption{The number of sinks formed by filament fragmentation, $N_\textsc{ff}$, and by disc fragmentation, $N_\textsc{df}$, in each simulation. Column 1 gives the random seed. Columns 2 \& 3 give the number of sinks formed by filament fragmentation and by disc fragmentation when $\delta_\textsc{sol}=0$. Similarly, columns 4 \& 5, 6 \& 7, 8 \& 9 and 10 \& 11  give the same quantities when, respectively, $\delta_\textsc{sol}=1/9$, $1/3$, $2/3$ and $1$.{The second from last row gives total number of stars over all random seeds. The last row gives the relative fraction of stars formed via filament and disc fragmentation.}}
\label{frag_table}
\end{table*}
}


\subsection{Influence of $\delta_\textsc{sol}$ on the stellar mass distribution}%

\begin{figure}
  \includegraphics[width=\columnwidth]{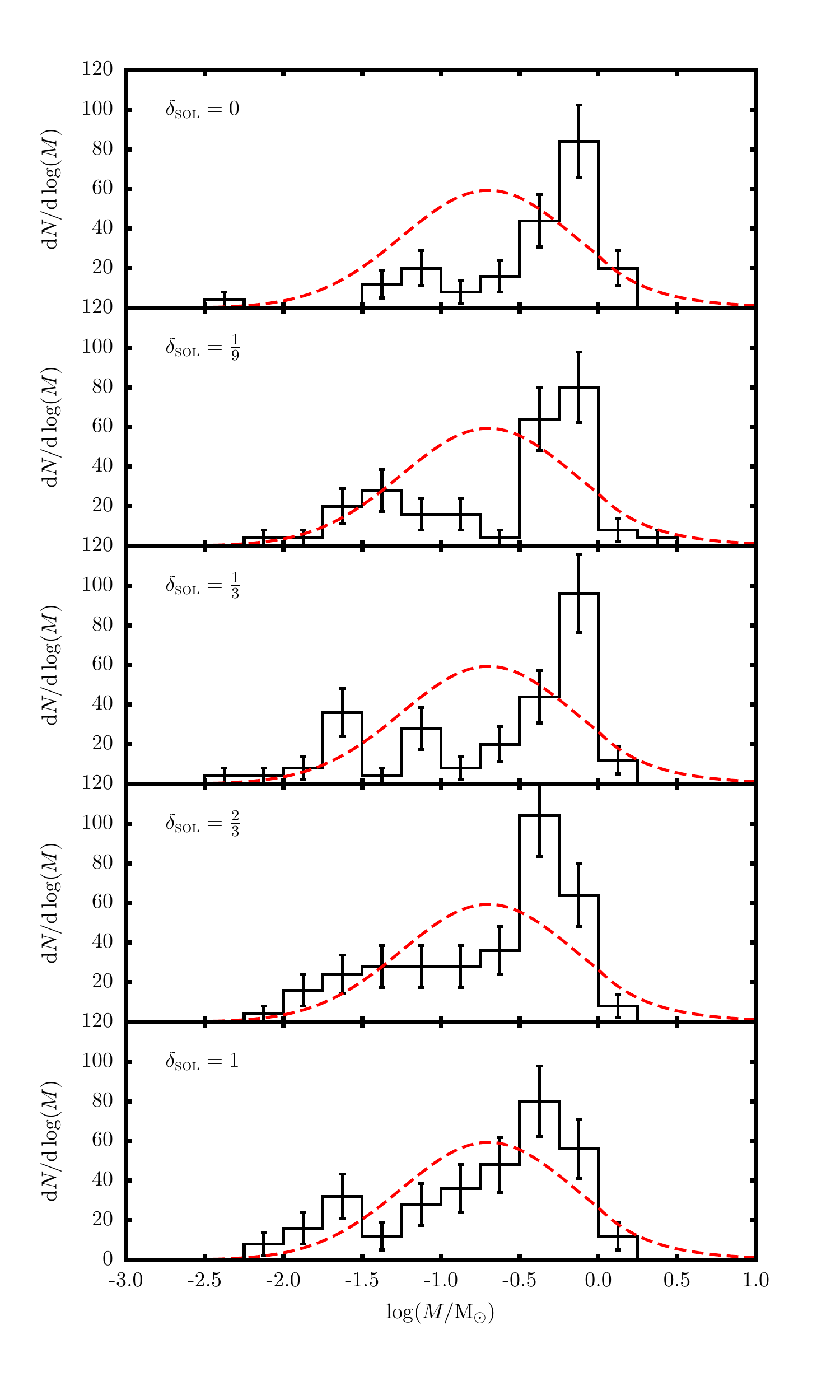}
  \caption{The black histograms show un-normalised stellar mass distributions, integrated over all values of $\mathcal{I}_\textsc{seed}$, for different values of $\delta_\textsc{sol}$. The error bars give the Poisson counting uncertainties. The red dashed lines show the \citet{C05} IMF, scaled to the area of the $\delta_\textsc{sol}=1$ histogram.}
  \label{turb_hist}
\end{figure}

\begin{figure}
  \includegraphics[width=\columnwidth]{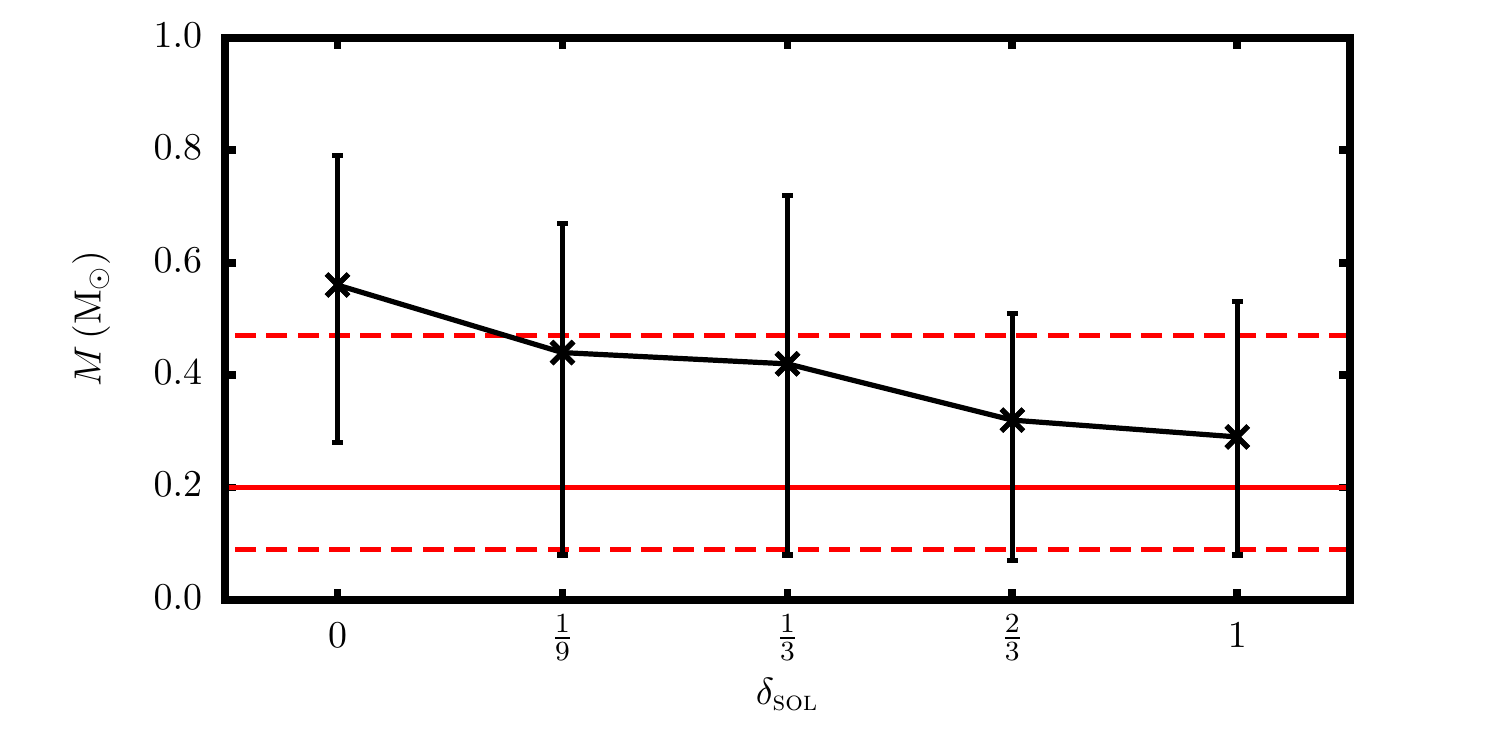}
  \caption{The black $\times$s give the median stellar mass, and the vertical black bars give the corresponding interquartile range, for different values of $\delta_\textsc{sol}$, integrated over all values of $\mathcal{I}_\textsc{seed}$. The solid and dashed horizontal red lines give the median and IQR for the \citet{C05} IMF.}
  \label{mass_median}
\end{figure}

Fig. \ref{turb_hist} shows the distribution of stellar masses formed with different values of $\delta_\textsc{sol}$, integrated over all values of $\mathcal{I}_\textsc{seed}$. Fig. \ref{mass_median} shows how the corresponding medians and interquartile ranges vary with $\delta_\textsc{sol}$. We see that, as $\delta_\textsc{sol}$ is increased, the median decreases monotonically, from $\sim\!0.6\,\mathrm{M}_\odot$ when $\delta_\textsc{sol}=0$, to $\sim\!0.3\,\mathrm{M}_\odot$ when $\delta_\textsc{sol}=1$. For reference, Fig. \ref{turb_hist} also shows the Chabrier (2005; hereafter C05) IMF (dashed red curve), and Fig. \ref{mass_median} shows the corresponding median and interquartile range (full and dashed horizontal red lines). However, we stress that we should not expect to reproduce the overall distribution of stellar masses observed in nature with simulated cores of a single mass, radius and non-thermal velocity dispersion, as treated here. We are simply seeking to establish what trends, if any, might derive from changing the mix of solenoidal and compressive modes in the imposed turbulent velocity field.

The decrease in median mass that results from increasing $\delta_\textsc{sol}$ can be attributed directly to the shift from filament fragmentation when $\delta_\textsc{sol}$ is low to disc fragmentation when $\delta_\textsc{sol}$ is high. When $\delta_\textsc{sol}$ is low, compressive turbulent modes create filaments, and these can be very effective at feeding matter from the periphery of the core, into the centre, where it forms a relatively massive star. Conversely, when $\delta_\textsc{sol}$ is high, solenoidal turbulent modes create discs around existing stars, and these discs tend to fragment to produce large numbers of low-mass companion stars.

\subsection{Multiplicity statistics}%

\subsubsection{Multiplicity frequency}\label{SEC:mfpf}%

Table \ref{mf_pf} lists the multiplicity frequencies and pairing factors extracted from the simulations. The procedures used to determine these statistics are detailed by \citet{LWHSW14b}. The multiplicity frequency, $mf$, is the fraction of systems which is multiple. The pairing factor, $pf$, is the mean number of orbits per system \citep[see][]{RZ93}. Thus, if $S$ is the number of single systems, $B$ the number of binaries, $T$, $Q$, etc. the numbers of triples, quadruples, etc., then
\begin{eqnarray}
mf&=&\frac{B+T+Q+...}{S+B+T+Q+...}\,,\\
pf&=&\frac{B+2T+3Q+...}{S+B+T+Q+...}\,
\end{eqnarray}
When all the multiple systems in a population are binary, $pf=mf$. When higher multiples are present (i.e. triples, quadruples, etc), $pf>mf$. We see from the table that there is no discernible trend in $mf$ with changing $\delta_\textsc{sol}$. Values of $mf$ range between 0.2 and 0.3, which is roughly the same as for M-dwarf stars in the field \citep[see][and references therein]{DK13}. In all cases, $pf>mf$, due to the presence of hierarchical multiple systems (up to sextuples). The variation of $pf$ is seemingly much more stochastic than that of $mf$. However, the stated uncertainties of $pf$ in Table \ref{mf_pf} are only lower limits.\footnote{$mf$ depends only on two numbers, $S$ and $B+T+Q+...=(N_\textsc{sys}-S)$. In contrast, $pf$ depends on $S$, $B$, $T$, $Q$, etc.; if we count up to sextuples, then six numbers. The higher the order of the system being counted, ${\cal O}$, the smaller the number of systems and hence the higher the fractional Poisson uncertainty. However, these systems are given higher weight in the numerator of $pf$, i.e. $w_{_{\cal O}}=({\cal O}-1)$, so their individual uncertainties are compounded in a way that they are not in $mf$.}

\begin{table}
\centering
\begin{tabular}{rccccc}\hline\\
  & $\delta_\textsc{sol}\!=\!0$ & $\delta_\textsc{sol}\!=\!\frac{1}{9}$ & $\delta_\textsc{sol}\!=\!\frac{1}{3}$ & $\delta_\textsc{sol}\!=\!\frac{2}{3}$ & $\delta_\textsc{sol}\!=\!1$ \\ \\ 
$N_\textsc{sys}$ & 32 & 34 & 51 & 61 & 55 \\
$mf$ & $0.28$ & $0.32$ & $0.20$ & $0.23$ & $0.29$ \\
$\Delta mf$ & $0.08$ & $0.08$ & $0.06$ & $0.05$ & $0.06$ \\
$pf$ & $0.63$ & $0.82$ & $0.29$ & $0.42$ & $0.49$ \\
$\Delta pf$ & $>0.18$ & $>0.20$ & $>0.08$ & $>0.10$ & $>0.10$ \\ \\ \hline
\end{tabular}
\caption{Multiplicity frequencies and pairing factors, for different values of $\delta_\textsc{sol}$, integrated over all values of $\mathcal{I}_\textsc{seed}$. Uncertainties on $mf$ are given by $\Delta mf=\sqrt{mf\,(1-mf)/(N_\textsc{sys})}$ where $N_\textsc{sys}$ is the total number of systems. Uncertainties on $pf$ are assumed to satisfy $\Delta pf>pf\,\Delta mf/mf$ (see footnote Section \ref{SEC:mfpf} for discussion).}
\label{mf_pf}
\end{table}

\subsubsection{Orbital properties}%

\begin{figure}
\includegraphics[width=\columnwidth]{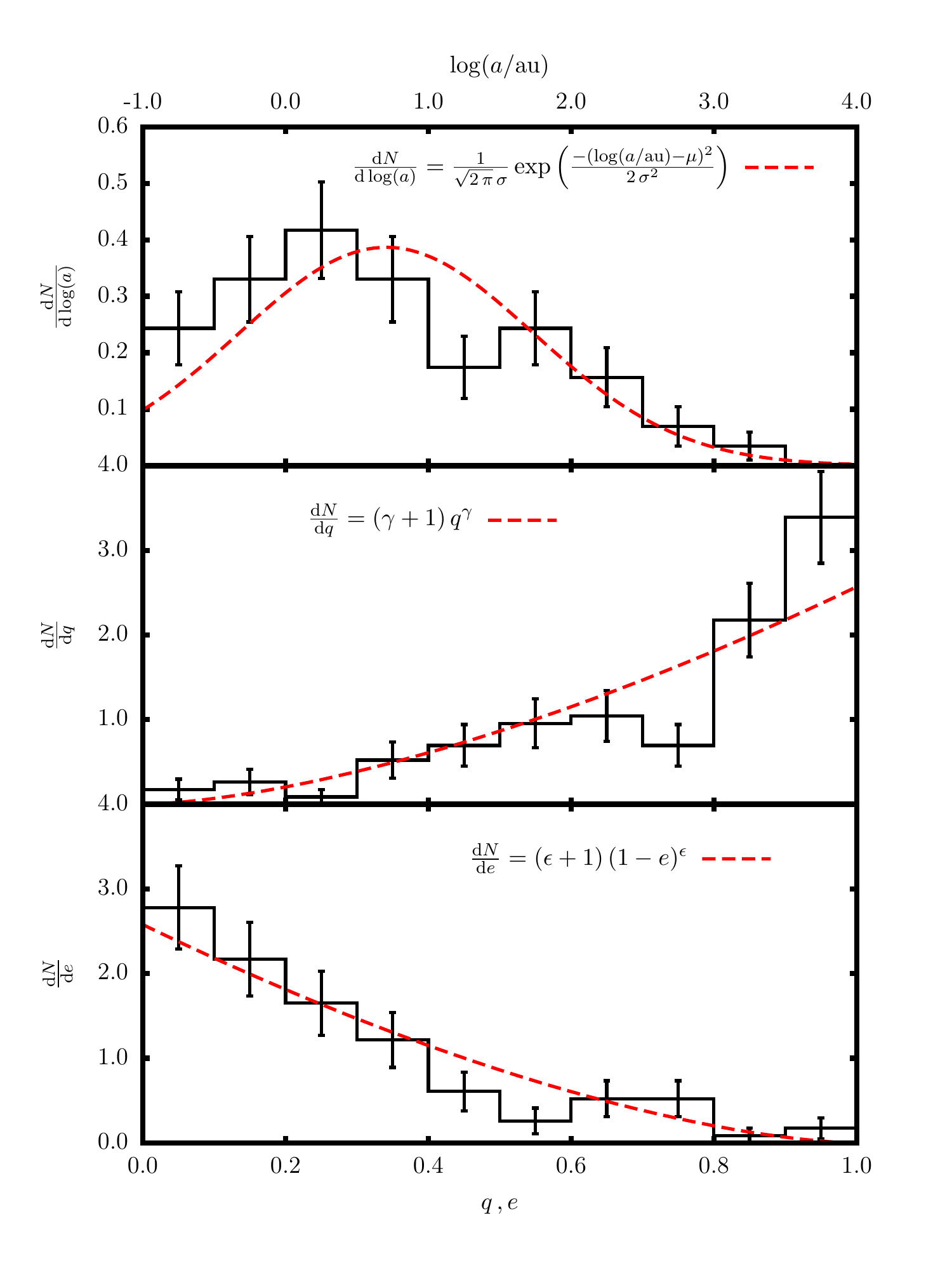}
\caption{Histograms showing the distribution of orbital properties integrated over all simulations (i.e. all $\delta_\textsc{sol}$ and $\mathcal{I}_\textsc{seed}$). The top panel shows the distribution of $\log_{_{10}}(a/{\rm au})$, where $a$ is the semimajor axis. The middle panel shows the distribution of mass ratios, $q$. The bottom panel shows the distribution of eccentricities, $e$. The red dashed lines are analytical fits to the data; fitting parameters are given in Table \ref{multi_params}.}
\label{multi_dist}
\end{figure}

The number of multiple systems formed in the simulations is too small to allow us to consider in detail how the distributions of orbital properties vary with $\delta_\textsc{sol}$. There is some tennuous evidence that, as $\delta_\textsc{sol}$ increases, the orbital eccentricities increase and the mass ratios decrease -- in other words, more compressive turbulence promotes more circular orbits and more closely matched companion masses -- but this is a very weak result. However, we can examine the combined distributions of semimajor axis $a$, mass ratio $q$ and eccentricity $e$. Fig. \ref{multi_dist} shows the distribution of $a$, $q$ and $e$ alongside analytic fits. The parameters of these fits are given in Table \ref{multi_params}. These distributions include all the orbits of hierarchical systems.

The semimajor axes range from $0.2\,\mathrm{au}$ to $2000\,\mathrm{au}$. The lower limit is set by the resolution of the simulations (i.e. the radius of a sink particle). The upper limit is compatible with observations of young embedded populations \citep[e.g. Taurus, Ophiuchus, etc,][]{KPPG12,KGPP12}.

If we fit the distribution of mass ratios with a power law of the form $\mathrm{d}N/\mathrm{d}q\propto q^\gamma$, we obtain $\gamma=1.6\pm0.2$, implying a strong preference for companions of comparable mass. Young embedded populations also show a preference for companions of comparable mass, but it is somewhat weaker, viz. $0.2\lesssim\gamma\lesssim1$ \citep{DK13}. To establish whether the observations could be fit more closely with a specific value of $\delta_\textsc{sol}$ would require a much larger ensemble of simulations.

If we fit the distribution of eccentricities with a reverse power law of the form $\mathrm{d}N/\mathrm{d}e\propto (1-e)^\epsilon$, we obtain $\epsilon=1.6\pm0.2$, implying a preference for low-eccentricity, similar to the preferentially circular orbits of field M-dwarfs \citep[see][Fig. 4]{DK13}.

\begin{table}
\centering
\begin{tabular}{cccc} \hline \\
$\mu_a$ & $\sigma_a$ & $\gamma$ & $\epsilon$ \\ \\
$0.7\pm0.1$ & $1.0\pm0.1$ & $1.6\pm0.2$ & $1.6\pm0.2$ \\ \\ \hline
\end{tabular}
\caption{Fitted multiplicity parameters integrated over all simulations (i.e. all $\delta_\textsc{sol}$ and $\mathcal{I}_\textsc{seed}$). $\mu_a$ and $\sigma_a$ are the mean and standard deviation of $\log_{_{10}}(a/{\rm au})$, where $a$ is the semimajor axis. $\gamma$ is the mass ratio distribution parameter $\mathrm{d}N/\mathrm{d}q\propto q^\gamma$. $\epsilon$ is the eccentricity distribution parameter $\mathrm{d}N/\mathrm{d}e\propto (1-e)^\epsilon$. $\gamma$ and $\epsilon$ are calculated using maximum likelihood estimation.}
\label{multi_params}
\end{table}

\section{Discussion and conclusions}\label{discussion}%

\subsection{Limitations}%

First, we stress that the work reported here is a numerical experiment, not an attempt to capture all the processes that ooccur in real prestellar cores. In particular we have simulated cores with a single mass, single radius, single non-thermal velocity dispersion, and single density profile. The only parameters varied are the fraction of turbulent energy that is solenoidal, $\delta_\textsc{sol}$, and the random seed that delivers different realisations, ${\cal I}_\textsc{seed}$. Therefore the results cannot be applied to all prestellar cores. Indeed, the results reported by LWH14 suggest that cores with masses $\ll1\,\mathrm{M}_\odot$ tend to form single stars, whilst cores with $\gtrsim1\,\mathrm{M}_\odot$ tend to form discs and filaments which fragment into multiple stars. The results presented here relate to the latter case.

Second, we have not included magnetic fields in the simulations. The three dimensional structure of a core's magnetic field is extremely difficult to extract from observations, and the evolution of the magnetic field requires the treatment of non-ideal MHD effects, along with the detailed ionisation chemistry of the matter. It is therefore beyond the scope of this investigation.

\subsection{Implications for star formation}%

$\delta_\textsc{sol}$ can only be determined observationally if one has the detailed three-dimensional velocity field, or if one has the detailed one-dimensional velocity field and a model, for example, assumed statistical isotropy of the velocity field \citep[see][]{BF14}. On the scale of cores one does not have this information (because of instrumental limitations, selection effects, and confusion), and one cannot make these assumptions.

The turbulent velocity field in a core is largely determined by the flows that create it. If a core is created by turbulent fragmentation \citep[e.g.][]{PN02,HC08,HC09} and there is a large inertial range between the scale at which the turbulent energy is injected (say by galactic shear) and the scale of the core, $\delta_\textsc{sol}$ is likely to approach its thermal value, i.e. $2/3$, and in this case disc fragmentation should be important. However, in reality there is probably not a clear division between the scale of energy injection and the inertial range leading to core formation; rather, turbulent energy is injected on many scales, including some that are smaller than, or comparable with, the core scale. In particular, when (if) a core condenses directly out of a shell swept up by an expanding nebula (H{\sc ii} region, stellar-wind bubble or supernova remnant), there may be a predominance of compressive turbulent energy, a very small inertial range, and hence a low value of $\delta_\textsc{sol}$, not so much disc fragmentation, and not so many low-mass stars.

\subsection{Comparison with similar work}

\citet{GFBK11} present adaptive mesh refinement simulations on scales intermediate between the molecular clouds simulated by, for example, \citet{BCB08}, and the cores treated in this work. They find that star formation occurs 25\% earlier (with respect to the start time of the simulation) when the velocity field is purely compressive, as opposed to solenoidal. They do not report an increase in the number of low mass stars or brown dwarfs formed when the turbulence is purely solenoidal. This may be because their minimum resolvable length scale is tens of $\mathrm{au}$, which is insufficient to capture the dynamics of disc fragmentation.


\subsection{Summary}%

We have shown that the collapse and fragmentation of a core is influenced by the fraction of turbulent energy that is solenoidal, $\delta_\textsc{sol}$, and hence -- by default -- by the fraction that is compressive. Specifically, as $\delta_\textsc{sol}$ is increased from $0$ (purely compressive) to $1$ (purely solenoidal), the proportion of stars that form by filament fragmentation decreases and the proportion that form by disc fragmentation increases; at the same time the number of stars formed increases, and their mean mass decreases. The formation of massive circumstellar discs requires $\delta_\textsc{sol}>1/3$. With the limited number of simulations that we have performed, we have been unable to establish any robust systematic trends in the multiplicity statistics that derive from varying $\delta_\textsc{sol}$.

\section*{Acknowledgements}%

OL and APW gratefully acknowledge the support of a consolidated grant (ST/K00926/1) from the UK STFC. {We also thank the considerate and constructive comments from the referee.} This work was performed using the computational facilities of the Advanced Research Computing @ Cardiff (ARCCA) Division, Cardiff University. All false-colour images have been rendered with \textsc{splash} \citep{P07}.

\bibliographystyle{mn2e}
\bibliography{refs}

\begin{thebibliography}{60}
\expandafter\ifx\csname natexlab\endcsname\relax\def\natexlab#1{#1}\fi

\bibitem[{{Alves}, {Lada} \& {Lada}(2001){Alves}, {Lada}, \& {Lada}}]{ALL01}
{Alves} J.~F., {Lada} C.~J., {Lada} E.~A., 2001, Nature, 409, 159

\bibitem[{{Andr{\'e}} {et~al}\mbox{.}(2007){Andr{\'e}}, {Belloche}, {Motte}, \&
  {Peretto}}]{ABMP07}
{Andr{\'e}} P., {Belloche} A., {Motte} F., {Peretto} N., 2007, A\&A, 472, 519

\bibitem[{{Bate}(1998)}]{B98}
{Bate} M.~R., 1998, ApJ, 508, L95

\bibitem[{{Bate}(2000)}]{B00}
{Bate} M.~R., 2000, MNRAS, 314, 33

\bibitem[{{Bate}(2009)}]{B09a}
{Bate} M.~R., 2009, MNRAS, 392, 590

\bibitem[{{Bate}(2012)}]{B12}
{Bate} M.~R., 2012, MNRAS, 419, 3115

\bibitem[{{Bate}(2014)}]{B14}
{Bate} M.~R., 2014, MNRAS, 442, 285

\bibitem[{{Bonnell}, {Clark} \& {Bate}(2008){Bonnell}, {Clark}, \&
  {Bate}}]{BCB08}
{Bonnell} I.~A., {Clark} P., {Bate} M.~R., 2008, MNRAS, 389, 1556

\bibitem[{{Brunt} \& {Federrath}(2014)}]{BF14}
{Brunt} C.~M., {Federrath} C., 2014, MNRAS, 442, 1451

\bibitem[{{Chabrier}(2003)}]{C03}
{Chabrier} G., 2003, ApJ, 586, L133

\bibitem[{{Chabrier}(2005)}]{C05}
{Chabrier} G., 2005, in Astrophysics and Space Science Library, Vol. 327, The
  Initial Mass Function 50 Years Later, {Corbelli} E., {Palla} F., {Zinnecker}
  H., eds., p.~41

\bibitem[{{Delgado-Donate}, {Clarke} \& {Bate}(2004){Delgado-Donate}, {Clarke},
  \& {Bate}}]{DCB04a}
{Delgado-Donate} E.~J., {Clarke} C.~J., {Bate} M.~R., 2004, MNRAS, 347, 759

\bibitem[{{Delgado-Donate} {et~al}\mbox{.}(2004){Delgado-Donate}, {Clarke},
  {Bate}, \& {Hodgkin}}]{DCB04b}
{Delgado-Donate} E.~J., {Clarke} C.~J., {Bate} M.~R., {Hodgkin} S.~T., 2004,
  MNRAS, 351, 617

\bibitem[{{Duch{\^e}ne} \& {Kraus}(2013)}]{DK13}
{Duch{\^e}ne} G., {Kraus} A., 2013, ARAA, 51, 269

\bibitem[{{Federrath}(2013)}]{F13}
{Federrath} C., 2013, MNRAS, 436, 1245

\bibitem[{{Federrath} {et~al}\mbox{.}(2010{\natexlab{a}}){Federrath},
  {Banerjee}, {Clark}, \& {Klessen}}]{FBCK10}
{Federrath} C., {Banerjee} R., {Clark} P.~C., {Klessen} R.~S.,
  2010{\natexlab{a}}, ApJ, 713, 269

\bibitem[{{Federrath} \& {Klessen}(2012)}]{FK12}
{Federrath} C., {Klessen} R.~S., 2012, ApJ, 761, 156

\bibitem[{{Federrath}, {Klessen} \& {Schmidt}(2008){Federrath}, {Klessen}, \&
  {Schmidt}}]{FKS08}
{Federrath} C., {Klessen} R.~S., {Schmidt} W., 2008, ApJ, 688, L79

\bibitem[{{Federrath} {et~al}\mbox{.}(2010{\natexlab{b}}){Federrath},
  {Roman-Duval}, {Klessen}, {Schmidt}, \& {Mac Low}}]{FRKS10}
{Federrath} C., {Roman-Duval} J., {Klessen} R.~S., {Schmidt} W., {Mac Low}
  M.-M., 2010{\natexlab{b}}, A\&A, 512, A81

\bibitem[{{Federrath} {et~al}\mbox{.}(2014){Federrath}, {Schr{\"o}n},
  {Banerjee}, \& {Klessen}}]{FSBK14}
{Federrath} C., {Schr{\"o}n} M., {Banerjee} R., {Klessen} R.~S., 2014, ApJ,
  790, 128

\bibitem[{Frigo \& Johnson(2005)}]{FFTW05}
Frigo M., Johnson S.~G., 2005, Proceedings of the IEEE, 93, 216, special issue
  on ``Program Generation, Optimization, and Platform Adaptation''

\bibitem[{{Gammie}(2001)}]{Gam01}
{Gammie} C.~F., 2001, ApJ, 553, 174

\bibitem[{{Girichidis} {et~al}\mbox{.}(2011){Girichidis}, {Federrath},
  {Banerjee}, \& {Klessen}}]{GFBK11}
{Girichidis} P., {Federrath} C., {Banerjee} R., {Klessen} R.~S., 2011, MNRAS,
  413, 2741

\bibitem[{{Goodwin} \& {Whitworth}(2004)}]{GW04}
{Goodwin} S.~P., {Whitworth} A.~P., 2004, A\&A, 413, 929

\bibitem[{{Goodwin}, {Whitworth} \& {Ward-Thompson}(2004){Goodwin},
  {Whitworth}, \& {Ward-Thompson}}]{GWW04}
{Goodwin} S.~P., {Whitworth} A.~P., {Ward-Thompson} D., 2004, A\&A, 423, 169

\bibitem[{{Goodwin}, {Whitworth} \& {Ward-Thompson}(2006){Goodwin},
  {Whitworth}, \& {Ward-Thompson}}]{GWW06}
{Goodwin} S.~P., {Whitworth} A.~P., {Ward-Thompson} D., 2006, A\&A, 452, 487

\bibitem[{{Harvey} {et~al}\mbox{.}(2001){Harvey}, {Wilner}, {Lada}, {Myers},
  {Alves}, \& {Chen}}]{HWL01}
{Harvey} D.~W.~A., {Wilner} D.~J., {Lada} C.~J., {Myers} P.~C., {Alves} J.~F.,
  {Chen} H., 2001, ApJ, 563, 903

\bibitem[{{Hennebelle} \& {Chabrier}(2008)}]{HC08}
{Hennebelle} P., {Chabrier} G., 2008, ApJ, 684, 395

\bibitem[{{Hennebelle} \& {Chabrier}(2009)}]{HC09}
{Hennebelle} P., {Chabrier} G., 2009, ApJ, 702, 1428

\bibitem[{{Horton}, {Bate} \& {Bonnell}(2001){Horton}, {Bate}, \&
  {Bonnell}}]{HBB01}
{Horton} A.~J., {Bate} M.~R., {Bonnell} I.~A., 2001, MNRAS, 321, 585

\bibitem[{{Hubber} {et~al}\mbox{.}(2011){Hubber}, {Batty}, {McLeod}, \&
  {Whitworth}}]{HBMW11}
{Hubber} D.~A., {Batty} C.~P., {McLeod} A., {Whitworth} A.~P., 2011, A\&A, 529,
  A27

\bibitem[{{Hubber}, {Walch} \& {Whitworth}(2013){Hubber}, {Walch}, \&
  {Whitworth}}]{HWW13}
{Hubber} D.~A., {Walch} S., {Whitworth} A.~P., 2013, MNRAS, 430, 3261

\bibitem[{{King} {et~al}\mbox{.}(2012{\natexlab{a}}){King}, {Goodwin},
  {Parker}, \& {Patience}}]{KGPP12}
{King} R.~R., {Goodwin} S.~P., {Parker} R.~J., {Patience} J.,
  2012{\natexlab{a}}, MNRAS, 427, 2636

\bibitem[{{King} {et~al}\mbox{.}(2012{\natexlab{b}}){King}, {Parker},
  {Patience}, \& {Goodwin}}]{KPPG12}
{King} R.~R., {Parker} R.~J., {Patience} J., {Goodwin} S.~P.,
  2012{\natexlab{b}}, MNRAS, 421, 2025

\bibitem[{{Kirk}, {Ward-Thompson} \& {Andr{\'e}}(2005){Kirk}, {Ward-Thompson},
  \& {Andr{\'e}}}]{KWA05}
{Kirk} J.~M., {Ward-Thompson} D., {Andr{\'e}} P., 2005, MNRAS, 360, 1506

\bibitem[{{Klessen}, {Heitsch} \& {Mac Low}(2000){Klessen}, {Heitsch}, \& {Mac
  Low}}]{KHM00}
{Klessen} R.~S., {Heitsch} F., {Mac Low} M.-M., 2000, ApJ, 535, 887

\bibitem[{{Kroupa}(2001)}]{K01}
{Kroupa} P., 2001, MNRAS, 322, 231

\bibitem[{{Lada} {et~al}\mbox{.}(2008){Lada}, {Muench}, {Rathborne}, {Alves},
  \& {Lombardi}}]{LMR08}
{Lada} C.~J., {Muench} A.~A., {Rathborne} J., {Alves} J.~F., {Lombardi} M.,
  2008, ApJ, 672, 410

\bibitem[{{Lomax} {et~al}\mbox{.}(2014){Lomax}, {Whitworth}, {Hubber},
  {Stamatellos}, \& {Walch}}]{LWHSW14}
{Lomax} O., {Whitworth} A.~P., {Hubber} D.~A., {Stamatellos} D., {Walch} S.,
  2014, MNRAS, 439, 3039

\bibitem[{{Lomax} {et~al}\mbox{.}(2015){Lomax}, {Whitworth}, {Hubber},
  {Stamatellos}, \& {Walch}}]{LWHSW14b}
{Lomax} O., {Whitworth} A.~P., {Hubber} D.~A., {Stamatellos} D., {Walch} S.,
  2015, MNRAS, 447, 1550

\bibitem[{{Matsumoto} \& {Hanawa}(2003)}]{MH03}
{Matsumoto} T., {Hanawa} T., 2003, ApJ, 595, 913

\bibitem[{{Matzner} \& {McKee}(2000)}]{MM00}
{Matzner} C.~D., {McKee} C.~F., 2000, ApJ, 545, 364

\bibitem[{{Morris} \& {Monaghan}(1997)}]{MM97}
{Morris} J.~P., {Monaghan} J.~J., 1997, Journal of Computational Physics, 136,
  41

\bibitem[{{Motte}, {Andre} \& {Neri}(1998){Motte}, {Andre}, \& {Neri}}]{MAN98}
{Motte} F., {Andre} P., {Neri} R., 1998, A\&A, 336, 150

\bibitem[{{Padoan} \& {Nordlund}(2002)}]{PN02}
{Padoan} P., {Nordlund} {\AA}., 2002, ApJ, 576, 870

\bibitem[{{Price}(2007)}]{P07}
{Price} D.~J., 2007, PASA, 24, 159

\bibitem[{{Reipurth} \& {Zinnecker}(1993)}]{RZ93}
{Reipurth} B., {Zinnecker} H., 1993, A\&A, 278, 81

\bibitem[{{Roy} {et~al}\mbox{.}(2014){Roy}, {Andr{\'e}}, {Palmeirim}, {Attard},
  {K{\"o}nyves}, {Schneider}, {Peretto}, {Men'shchikov}, {Ward-Thompson},
  {Kirk}, {Griffin}, {Marsh}, {Abergel}, {Arzoumanian}, {Benedettini}, {Hill},
  {Motte}, {Nguyen Luong}, {Pezzuto}, {Rivera-Ingraham}, {Roussel}, {Rygl},
  {Spinoglio}, {Stamatellos}, \& {White}}]{RAP13}
{Roy} A. {et~al.}, 2014, A\&A, 562, A138

\bibitem[{{Schmidt} {et~al}\mbox{.}(2009){Schmidt}, {Federrath}, {Hupp},
  {Kern}, \& {Niemeyer}}]{SFHK09}
{Schmidt} W., {Federrath} C., {Hupp} M., {Kern} S., {Niemeyer} J.~C., 2009,
  A\&A, 494, 127

\bibitem[{{Stamatellos} {et~al}\mbox{.}(2011){Stamatellos}, {Maury},
  {Whitworth}, \& {Andr{\'e}}}]{SMWA11}
{Stamatellos} D., {Maury} A., {Whitworth} A., {Andr{\'e}} P., 2011, MNRAS, 413,
  1787

\bibitem[{{Stamatellos} \& {Whitworth}(2008)}]{SW08}
{Stamatellos} D., {Whitworth} A.~P., 2008, A\&A, 480, 879

\bibitem[{{Stamatellos} \& {Whitworth}(2009{\natexlab{a}})}]{SW09a}
{Stamatellos} D., {Whitworth} A.~P., 2009{\natexlab{a}}, MNRAS, 392, 413

\bibitem[{{Stamatellos} \& {Whitworth}(2009{\natexlab{b}})}]{SW09b}
{Stamatellos} D., {Whitworth} A.~P., 2009{\natexlab{b}}, MNRAS, 400, 1563

\bibitem[{{Stamatellos} {et~al}\mbox{.}(2007){Stamatellos}, {Whitworth},
  {Bisbas}, \& {Goodwin}}]{SWBG07}
{Stamatellos} D., {Whitworth} A.~P., {Bisbas} T., {Goodwin} S., 2007, A\&A,
  475, 37

\bibitem[{{Stamatellos}, {Whitworth} \& {Hubber}(2011){Stamatellos},
  {Whitworth}, \& {Hubber}}]{SWH11}
{Stamatellos} D., {Whitworth} A.~P., {Hubber} D.~A., 2011, ApJ, 730, 32

\bibitem[{{Stamatellos}, {Whitworth} \& {Hubber}(2012){Stamatellos},
  {Whitworth}, \& {Hubber}}]{SWH12}
{Stamatellos} D., {Whitworth} A.~P., {Hubber} D.~A., 2012, MNRAS, 427, 1182

\bibitem[{{Toomre}(1964)}]{T64}
{Toomre} A., 1964, ApJ, 139, 1217

\bibitem[{{Walch} {et~al}\mbox{.}(2009){Walch}, {Burkert}, {Whitworth}, {Naab},
  \& {Gritschneder}}]{WBWNG09}
{Walch} S., {Burkert} A., {Whitworth} A., {Naab} T., {Gritschneder} M., 2009,
  MNRAS, 400, 13

\bibitem[{{Walch} {et~al}\mbox{.}(2010){Walch}, {Naab}, {Whitworth}, {Burkert},
  \& {Gritschneder}}]{WNWB10}
{Walch} S., {Naab} T., {Whitworth} A., {Burkert} A., {Gritschneder} M., 2010,
  MNRAS, 402, 2253

\bibitem[{{Walch}, {Whitworth} \& {Girichidis}(2012){Walch}, {Whitworth}, \&
  {Girichidis}}]{WWG12}
{Walch} S., {Whitworth} A.~P., {Girichidis} P., 2012, MNRAS, 419, 760

\end{thebibliography}

\label{lastpage}
\end{document}